\documentclass[aps,pre,twocolumn,showpacs,superscriptaddress,groupedaddress]{revtex4}

\usepackage{amsmath}
\usepackage{graphicx}
\usepackage{amssymb}
\usepackage{bm}
\usepackage{longtable}
\usepackage{color}

\usepackage{times}
\usepackage{hyperref}
\usepackage[T1]{fontenc}
\usepackage{textcomp}
\usepackage{mathptmx}
\usepackage{epstopdf}

\newcommand{\V}[1]{\boldsymbol{ #1}}
\newcommand{\mean}[1]{\left\langle {#1} \right\rangle }
\newcommand{\paren}[1]{\left( {#1} \right ) }

\newcommand{\caja}[1]{\left[ {#1} \right ] }

\newcommand{\E}{\mathrm{e}}
\newcommand{\D}{\mathrm{d}}

\begin{document}

\title{Liquid-glass transition in equilibrium}

\author{G. Parisi}
\affiliation{Dipartimento di Fisica, Sapienza Universit\`a di Roma, Piazzale A. Moro 2, I-00185, Rome, Italy}
\affiliation{INFN, Sezione di Roma I, IPFC-CNR, P.le A. Moro 2, I-00185 Roma, Italy}
\author{B. Seoane}
\affiliation{Dipartimento di Fisica, Sapienza Universit\`a di Roma}
\affiliation{Instituto de Biocomputaci\'on and
  F\'{\i}sica de Sistemas Complejos (BIFI), 50009 Zaragoza, Spain.}

\begin{abstract}

  We show in numerical simulations that a system of two coupled
  replicas of a binary mixture of hard spheres undergoes a phase transition
  in equilibrium at a density slightly smaller than the glass transition
  density for an unreplicated system. This result is in agreement with
  the theories that predict that such a transition is a precursor of
  the standard ideal glass transition. The critical properties are
  compatible with those of an Ising system. The relations of this
  approach to the conventional approach based on configurational
  entropy are briefly discussed.

\end{abstract}

\maketitle Glass forming materials display a rapid growth of the
viscosity upon cooling~\cite{ediger:96}. Dynamics is dramatically
slowed down, but this fact is not accompanied by any obvious
structural or thermodynamic change~\cite{leheny:96}. As a consequence,
below certain temperature, the liquid gets trapped in a solid like
amorphous configuration for a very long time. From the experimental
point of view, these systems live permanently out of equilibrium: it
is natural to ask whether this phenomenon is a consequence of a
thermodynamic transition or, in contrast, whether it is just a pure
dynamical arrest process~\cite{berthier:11}.  Different mean field
approaches, from the Adam-Gibbs theory~\cite{adam:65}, to the mode
coupling theory~\cite{gotze:92} or the spin-glass
theory~\cite{mezard:87} agree on the existence of an ``ideal
structural glass transition'' in the infinite time limit, but the
validity of this claim for realistic systems is still under debate;
other interpretations of the phenomenon where no transition is present
have been proposed~\cite{chandler:10}.

Under the mean field approximations, supercooled liquids undergo a random
first order transition
(RFOT)~\cite{kirkpatrick:87,kirkpatrick:89}, which corresponds to
``one-step replica symmetry breaking''~\cite{mezard:87}.  In this scheme,
below certain temperature $T_\mathrm{c}$, the ergodicity is lost due to the
appearance of an exponentially large number of metastable states. The system
gets trapped in one of them (not necessarily the thermodynamic) and the
large relaxation times are thus related to the escape times. The Kauzmann-like
entropy crisis may take  place at  $T_\mathrm{K}(<T_\mathrm{c})$, the point where
the ideal glass phase becomes the thermodynamic one.

The properties of this transition can be studied (at fixed density
$\rho$) by considering the replica potential $W(q)$ (i.e., the free
energy) as a function of the degree of similarity between all the
possible amorphous configurations~\cite{franz:97}. In analogy with
spin glasses, the chosen order parameter is the ``overlap'' ($q$)
between the configurations of two equilibrium systems (``replicas'').
A key prediction of the theory is that one can observe a precursor of
the phase transition in the shape of this potential still deep in the
liquid phase.  Indeed, the glass transition is characterized by a
sharp decrease of the number of available states, which should be
detected by the appearance of second minima in $W(q)$ at large $q$. In
contrast to ordinary first order transitions, these two minima are not
related to different phases but to similar and completely different
configurations.  In the RFOT approach this transition should survive
in the limit of zero coupling: in other approaches this transition may
exist at non-zero coupling, but it would disappear at zero coupling as
stressed by~\cite{garrahan:13}.

Since $T_\mathrm{K}$ is well below $T_\mathrm{c}$, detecting directly the
two-well structure in $W(q)$ in a numerical simulation is very difficult in
practice. However, the situation improves if one adds an external field
$\epsilon$ conjugate to $q$ that couples the two replicas,
\begin{equation} H_\mathrm{tot}(\V{R}_1,\V{R}_2)=
  H(\V{R}_1)+H(\V{R}_2)-\epsilon\, q(\V{R}_1,\V{R}_2), \end{equation}
 $\V{R}_\alpha$ being the shorthand for the whole set of particle
positions in replica $\alpha$, $H_\mathrm{tot}$ the total Hamiltonian,
and $H$ the internal interaction at each replica. The free energy
 density $f(\epsilon)\equiv F(\epsilon)/N$ is given by
\begin{equation}
f(\epsilon)=\min_  q W(q)-\epsilon q \ .
\end{equation}
In the presence of this external field, the glass transition point
becomes a coexistence line $\epsilon(T)$ separating the low and
high $q$ regions, that extends to higher temperatures,
terminating in a critical point (exactly as in the more mundane
gas-liquid transition).  Strong evidences for the existence of such a
coexistence line beyond mean field have been presented
recently~\cite{cammarota:10,berthier:13}. 

We are interested here in studying this critical point where the first order
line appears in a system of hard spheres. In this case, the density (not the
temperature) will play the role of control parameter. At least in mean field,
where the replica potential $W(q)$ is an analytic function of $q$ and of the
other parameters (such as density or temperature), the critical point is fixed
by the condition
 \begin{eqnarray}
 W''(q^*)=0,\\W'''(q^*)=0 \, ,
 \end{eqnarray}
 which is equivalent to 
 \begin{eqnarray}
 W(q)=W(q^*)+W'(q^*)(q-q*)+\frac{g}{4} (q-q*)^4\nonumber\\+O\paren{(q-q*)^5} \, .
 \end{eqnarray}
 
 At first sight the physics looks very similar to a gas liquid
 transition and thus, should be in the same universality class of the
 Ising ferromagnetic phase transition. However a more careful analysis
 shows that the situation is more subtle and crucially depends on the
 details.  As shown in \cite{franz:98-1} we can introduce two slightly
 different potentials: the quenched potential, where the field
 $\epsilon$ acts only on one of the replicas and the annealed
 potential, where the field $\epsilon$ acts on both replicas. It can
 be shown that the first case is in the universality class of the
 ferromagnetic Ising model with a random quenched random magnetic
 field (see also~\cite{franz:13,biroli:13}), while the second case
 corresponds to a pure Ising case. We expect that the second case
 should be much easier to simulate since the random ferromagnetic
 Ising model approaches equilibrium very slowly. In this paper we
 consider this second case (the annealed one), which is not common in
 numerical studies (in particular is different to the one considered
 in \cite{cammarota:10}), but has been recently studied in
 \cite{berthier:13}.

 The study of $W(q)$ has traditionally been inaccessible for computer
 simulations. Indeed, in practice, the two separate configurations
 decorrelate quickly, which leaves little time to sample the high
 overlap region of the probability distribution function
 $p(q)$. However, in the last years, constrained Monte Carlo (MC)
 methods have been proposed as a solution to compute this
 $W(q)$~\cite{cammarota:10, berthier:13}. Here we propose a recent
 constrained MC method, the tethered method~\footnote{This method is
   not related to the tether method~\cite{speedy:93}.}, originally proposed for spin lattice
 systems~\cite{martinmayor:07,fernandez:09,martinmayor:11} but recently
 applied to hard spheres~\cite{fernndez:12}.  This method presents a major
 simplification of standard umbrella sampling
 method~\cite{torrie:74,torrie:77, bartels:00} since the potential differences
 are very precisely computed from a thermodynamic integration, thus avoiding
 the tedious multi histogram reweightings.

We study the model introduced in~\cite{brambilla:09}: a $50:50$ binary
mixture of hard spheres (HSs) where the diameter of the larger
particle, $d_B$, is $1.4$ times the diameter $d_A$ of the smaller
one. This high dispersion between particles sizes prevents the
crystallization. We study systems of $N=62,\ 124,\ 250,\text{ and } 500$
particles. In addition, all the simulations reported here are
performed at constant volume, parametrized through the volume fractions
$\phi=\pi N \paren{d_A^3+d_B^3}/12V$. The
simulation box is cubic, $V=L^3$ with periodic boundary
conditions.

As discussed before, we are interested in studying the degree of
similarity between different configurations as a function of $\phi$. For
this reason, in the following, we will consider simultaneously two
copies of the system labeled $\alpha=1,2$. The distance between these
two configurations can be measured with the overlap $q_{1,2}$.
There are two possible definitions. The first one, introduced in
\cite{franz:98-1,mzard:99}, is
\begin{equation}
q_{1,2}=\frac1{N} \sum_i\  v\paren{|\V{r}^{(1)}_i-\V{r}^{(2)}_i|}\, ,
\end{equation}
where $\V{r}^{(\alpha)}_i$ represents the position of the $i$th
particle in the replica $\alpha$, and $v$ is a function that is 1 at short
distances and goes very fast to $0$ at distances greater than some
fraction of the interparticle distance.

Here, for practical reasons, we prefer to use a different but very similar,
definition of the overlap, that is the same used in
\cite{cammarota:10} [Ref. \cite{berthier:13} uses the previous definition
of the overlap, following verbatim \cite{cardenas:99},
i.e., $v(r)=\theta(r-a)$ using the same value of $a=0.3$]. We divide
our simulation box into $N_c$ small cubic boxes. To each box $i$ in
the replica $\alpha$, we assign an occupation variable
$n_{i,\;T}^{(\alpha)}=1$ in the case where it contains a particle of type
$T(=A,B)$, and $n_{i,\;T}^{(\alpha)}=0$ if it does not. The linear
size of the cell, $\ell$, is chosen to guarantee that two different
particles can never occupy the same cell. This condition is fulfilled
by taking the largest possible number of cells, $N_\mathrm{c}$,
compatible with the constraint $\ell<d_A/\sqrt{3}$ (i.e. the largest
diagonal of the cube is smaller than $d_A$)~\footnote{In the range of
  volumes simulated $0.548\le\ell\le 0.574$.}. Our overlap is then
defined as
\begin{equation}\label{eq:overlap}
q_{1,2}=\frac{1}{N_\mathrm{c}}\sum_{i=1}^{N_\mathrm{c}}n_{i,\;A}^{(1)}\,n_{i,\;A}^{(2)}+n_{i,\;B}^{(1)}\,n_{i,\;B}^{(2)}.
\end{equation}
Within this definition, the overlap between two identical
configurations is $q_{1,2}=1$, while for two completely uncorrelated
configurations it is $q_{1,2}=q_0=N/2N_\mathrm{c}$~\footnote{Due to the
  integer nature of $N_\mathrm{c}$, the actual value of $q_0$ changes
  with $\phi$: $q_0=0.027(\phi=0.3)$, $0.035(\phi=0.4)$,
  $0.037 (0.45<\phi<0.52)$ and $0.051 (0.52\le\phi\le 0.57)$.}.

The free energy cost of maintaining the two thermalized replicas
of the system at a given $q$ is
\begin{equation}\label{eq:Wq}
W(q)=-\frac1{N}\,\log{\int\int\mathrm{d} \V{R}_1\mathrm{d} \V{R}_2\  \mathcal{H}(\V{R}_1)\mathcal{H}(\V{R}_2)\ \delta\paren{q-q_{1,2}}}\,,
\end{equation} where
$\mathcal{H}(\V{R}_\alpha)=0$ if any pair of spheres in the replica
$i$ overlaps, or 1 otherwise ($\V{R}_\alpha$ being the abbreviation
for $\{\V{r}_i ^{(\alpha)}\}_{i=1}^N$, the set of all the $N$ particle
positions in the replica $\alpha$).  We propose a slight variation of
this last definition. Instead, we consider its convolution with a
strongly peaked Gaussian centered on $q$ with variance $(k
N)^{-1}$~\footnote{For the convolution, one needs a Gaussian narrow
  enough to be able to separate the low and the high $q$ peaks, but
  wide enough to let the simulation evolve. Since the amount of
  particles we are considering here is quite small, we need to add an
  extra factor $k$ to the variance $N^{-1}$ (otherwise the tethered
  simulation was not able to constrain the value of $q$). Here we used
  $k=200$, as was chosen in a previous work in hard
  spheres~\cite{fernndez:12}. This tunable parameter $k$ does not
  modify the mean value of $q$, but is present as a constant
  multiplicative factor in the definition of the replica
  field~\eqref{eq:replica-field}. The real probability distribution
  function for the overlap (without the convolution) is recovered by a
  thermodynamic integration of this field, which means that the actual
  value of $k$ only contributes to form of its normalization constant,
  which has no effect on the physical properties of the system.}:
\begin{equation}\label{eq:hatWq}
  \hat{W}(q)=-\frac1{N}\,\log{\int\int\mathrm{d} \V{R}_1\mathrm{d} \V{R}_2\  \mathcal{H}(\V{R}_1)\mathcal{H}(\V{R}_2)\,\E^{-\frac{k N}{2}{}\paren{q-q_{1,2}}^2}}\,.
\end{equation}
These two definitions are equivalent in the thermodynamic
limit, but this latter is better from a MC simulation point of view.
To explain why, let us take the derivative of Eq.~\eqref{eq:hatWq} with respect to $q$:
\begin{equation}\label{eq:derhatWq}
  \hat{W}'(q)=\frac{\int\int\mathrm{d} \V{R}_1\mathrm{d} \V{R}_2\ k \caja{q-q_{1,2}}\omega_N (\V{R}_1,\V{R}_2,V;q)}{\int\int\mathrm{d} \V{R}_1\mathrm{d} \V{R}_2\ \omega_N (\V{R}_1,\V{R}_2,V;q) },
\end{equation}
with
\begin{eqnarray}\label{eq:weight}
 \omega_N (\V{R}_1,\V{R}_2,\phi;q)=\mathcal{H}(\V{R}_1)\mathcal{H}(\V{R}_2)\, \E^{-\frac{k N}{2} \caja{q -
    q_{1,2}(\V{R}_1,\V{R}_2)}^2}.
\end{eqnarray}
That means that the {\em replica field} can be understood as the MC thermal
average obtained with the {\em tethered} measure \eqref{eq:weight},
 \begin{equation}\hat W'(q)=\mean{\hat\epsilon}_q,\ \hat\epsilon=k \paren{q-q_{1,2}}.\label{eq:replica-field}\end{equation}
 With this idea, we build a new ensemble, where not only the volume
 and number of particles are fixed, but there is also an additional
 soft constraint for the averaged overlap $\mean{q_{1,2}}_q\approx q$.
 Once the field is obtained as function of $q$, the replica potential
 $W(q)$ can be easily computed by a thermodynamic integration.  It is
 interesting to point out that although the new weight is formally
 equal to that of traditional umbrella sampling (and so are the
 simulations), we skip the step of reconstructing the unbiased
 probability distribution function of $q$ and its tedious
 multi histogram reweightings (see ~\cite{berthier:13} for an example
 of this procedure in a similar problem). Indeed, in the umbrella
 sampling approach, one needs to compute $p(q)$ to obtain the overlap
 potential [$W(q)=-\frac1{N}\log P(q)$], which is much costlier in time and
 less precise than only recording the central point of the distribution
 and performing a line integral with it.

 We run simulations at $0.3\le\phi\le 0.57$ at $21$ values of $q$
 evenly spaced between 0 and 1. In addition, we consider five (100 for
 $N=500$ and 50 for $N=250$ for $0.54\le \phi\le 0.57$) realizations
 of each experiment, and results presented here are averaged over all
 these samples.  The set up is the following. We start with a
 thermalization of each of the two replicas. We consider
 thermalizations of $\tau_0$ elementary MC steps (EMCSs)\footnote{ Our
   thermalization times were $\tau_0=10^7$ for $0.3\le\phi\le 0.45$,
   $\tau_0=10^8$ for $0.5\le\phi\le 0.54$, and $\tau_0=5\times 10^8$
   for $0.55\le\phi\le 0.57$.  }, defining EMCS as $N$ attempts at
 ordinary individual random particle moves. Only once the initial
 configurations are thermalized, we run the tethered simulations at
 fixed $q$ using the weight~\eqref{eq:weight}. In order to ensure the
 thermalization, we consider two alternative experiments. On the one
 hand, we move sequentially from $q=1$ to $0$ in steps of $0.05$, and
 on the other, we consider the reverse procedure.  At each value of
 $q$ we remain $\tau_\mathrm{int}=0.1\tau_0$ EMCS, and using the
 latter $0.05\tau_0$ EMCS in the analysis. We completely avoid
 hysteresis effects for $\phi\le 0.57$.  We have systematically
 checked that both cycles are compatible, but the results presented in
 this paper correspond only to the $q$-descending cycle.

\begin{figure}[h]
\includegraphics[angle=270,width=\columnwidth]{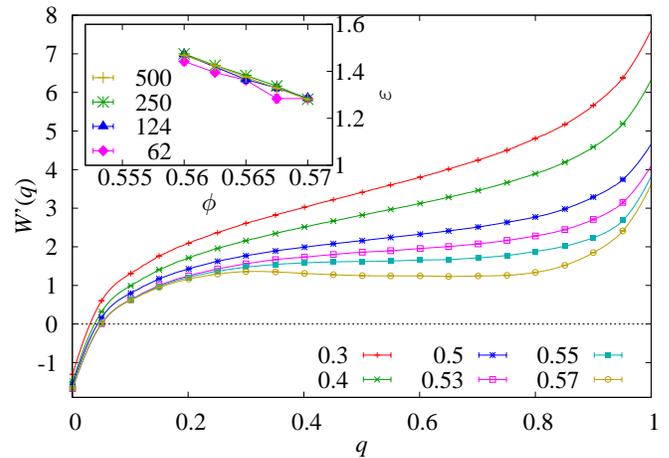}
\caption{(Color online) Derivative of the replica potential, $\hat W'(q)$
  obtained as ~\eqref{eq:hatWq} at different packing fractions. Inset:
  Phase diagram $(\epsilon,\phi)$ extracted from the Maxwell
  construction for different system sizes. The errors are computed using
  the jackknife method. }
\label{fig:field_and_phasediag}
\end{figure}

The generalization of this formalism to the presence of an external
field $\epsilon$ coupling the two replicas is straightforward.
Indeed, the probability distribution density for $q$ at a given $\epsilon$
is just the free one, multiplied by a constant exponential factor,
i.e.,  $P_\epsilon (q)\propto \exp\caja{-\paren{N  W(q)-\epsilon
    q}}$. Furthermore~\cite{martinmayor:07}, 
\begin{equation}\label{eq:diffprob}
  \log \hat P_\epsilon (q_2)-\log \hat P_\epsilon (q_1)=N\int_{q_1}^{q_2}\D q \caja{ \mean{\hat \epsilon}_q-\epsilon},
\end{equation}
with $ \mean{\hat \epsilon}_q=\hat W'(q)$ given by~\eqref{eq:hatWq}. Thus,
one just needs to simulate the $\epsilon=0$ case, and the results for
$\epsilon>0$ are obtained by displacing $\mean{\hat \epsilon}_q$
precisely by $\epsilon$.  We display in the main panel of
Fig.~\ref{fig:field_and_phasediag} the replica field $W'(q)$
obtained at different values of $\phi$. In the thermodynamically
stable region $\hat W'(q)$ is monotonically growing from zero, and the
equilibrium state (the maximum of $P_\epsilon$) is given by the single
root $\hat W'(q)=\epsilon$. The situation is rather different when
metastability begins. In finite systems, the phase separation has the
direct consequence of the apparition of spinoidals in $\hat W'(q)$
corresponding to the low and high overlap regions. The coexistence
condition $P_{\epsilon_\text{co}}(q_{\text{
    low}})=P_{\epsilon_\text{co}}(q_{\text{high}})$, as can be
directly obtained from \eqref{eq:diffprob}, is equivalent to a Maxwell
construction. We show the inset of
Fig.~\ref{fig:field_and_phasediag} the phase diagram obtained with
this process.

\begin{figure}[h]
\includegraphics[angle=270,width=\columnwidth]{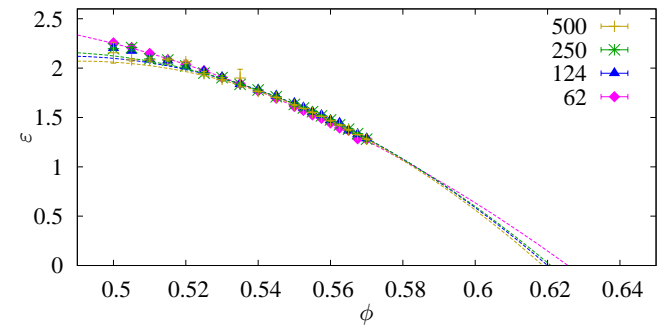}
\caption{(Color online) Values of $\epsilon$ that guarantee a balanced
  distribution of $P_\epsilon(q)$, as function of $\phi$. We also
  included the fits of these curves to a second order polynomial in
  $\phi$. The extrapolations of $\phi_\mathrm{K}$ using these fits are
  collected in Table~\ref{tab:fits}.}
\label{fig:epsilon}
\end{figure}

As was discussed before, the first order transition line is expected
to extend within the liquid phase in the presence of an external field
but terminates in a critical point. This point cannot be detected by
the Maxwell construction so we propose a different approach. Besides,
the coexistence line can be extended beyond the critical point by what
is known as the Widom line~\cite{widom:65} that is characterized by
$W'''(q)=0$. With this idea in mind, we look for the $\epsilon$ value
that makes the distribution $P_\epsilon(q)$ balanced. In particular,
we seek the $\epsilon$ that causes the skewness to vanish.  In the
metastability region, the $\epsilon(\phi)$ line obtained with this
method (see Fig.~\ref{fig:epsilon}) is numerically indistinguishable
from the one computed using the Maxwell construction (as we would
expect in the infinite volume limit). Indeed, it finds the field at
which the probability distribution function has two peaks of equal
probability. Once it is understood that this line contains the second order
transition point, we can try to infer its location by seeking
universal behavior. At least in mean field, this point should belong
to the $d=3$ Ising model universality class~\cite{franz:98-1}, which
means that its critical exponents are known~\cite{pelissetto:02}, thus
making easier the computation of $\phi_\mathrm{c}$.

\begin{figure}[h]
\includegraphics[angle=270,width=\columnwidth]{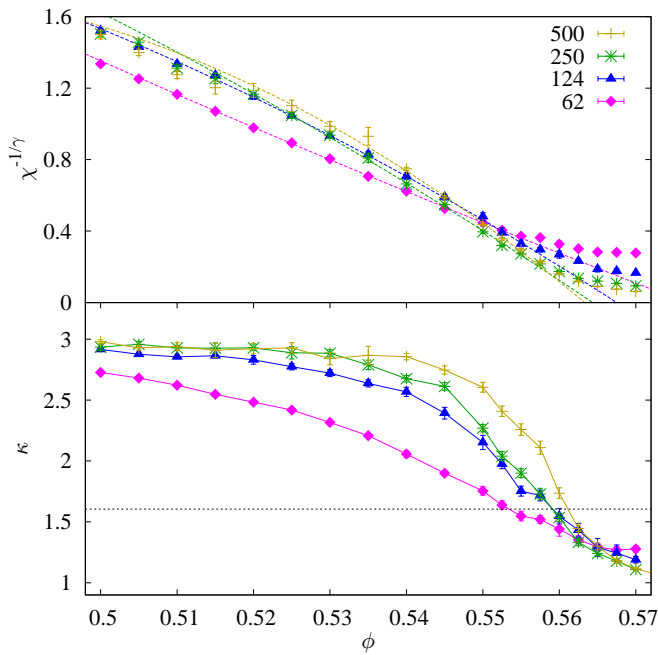}
\caption{(Color online) Top: static susceptibility at different $N$
  scaled as $\chi^{-1/\gamma}$ with $\gamma=1.2372$. We include the
  extrapolations of $\phi_\mathrm{c}$. Bottom: kurtosis at different
  $N$, the dotted line indicates the universal value for the kurtosis $\kappa=1.6043(10)$ at the critical point~\cite{blote:95}.}
\label{fig:chi-binder}
\end{figure}

We start with the static susceptibility
$\chi=N[\mean{q^2}-\mean{q}^2]$. As usual in the vicinity of a second
order transition, it should scale as
$\chi\propto|\phi-\phi_\mathrm{c}|^{-\gamma}$, in this case with
$\gamma=1.2372$. As can be seen in Fig.
~\ref{fig:chi-binder} (top) we find a collapse of the data at
different system sizes below $\phi_\mathrm{c}$ using this scaling (at
least for $N\ge 124$).  We obtain $\phi_\mathrm{c}^N$ from an
 extrapolation to a second order polynomial (see Table~\ref{tab:fits} for the fitting
details). These values agree with the area of $\phi$ 
where the kurtosis of the distribution,
\begin{eqnarray}
\kappa=\frac{\mean{m}^4}{\mean{m^2}^2}&\text{ with }&m=q-\mean{q},
\end{eqnarray}
intersects for increasing $N$, as shown in
Fig.~\ref{fig:chi-binder} (bottom). Indeed, the kurtosis is related to the
Binder cumulant~\cite{binder:82}, by $B=1-\kappa/3$.  Like this
cumulant, the kurtosis is universal at the critical point, and its
value is known $\kappa=1.6043(10)$~\cite{blote:95}. Within the
precision, we can say that the data are compatible with both statements.

\begin{table}[b]
\centering
\begin{tabular*}{\columnwidth}{@{\extracolsep{\fill}}  c | c  c | c  c }
  \hline \hline
  $N$ & $\phi^N_\mathrm{c}$ & $\chi^2/\mathrm{dof}$ & $\phi^N_\mathrm{K}$ & $\chi^2/\mathrm{dof}$ \\
  \hline 
62         & 0.5766(3)(15)   & 2.5/6   & 0.626(2) & 12/11\\
124        & 0.567(3)(4)   & 1.7/6   & 0.620(2)   & 19.5/11\\
250        & 0.5644(13)(16)   & 8.5/6   & 0.6209(12)   & 8.8/11\\
500        & 0.5633(6)(17)   & 8.5/6   & 0.6187(7)   & 5.7/11\\\hline \hline 
\end{tabular*}
\caption{Extrapolations for $\phi_\mathrm{c}$ and $\phi_\mathrm{K}$
  obtained from the fits of $\chi$ (displayed in
  Fig.~\ref{fig:chi-binder}) and $\epsilon(\phi)$ (in
  Fig.~\ref{fig:epsilon}). In both cases, data at each $N$ are
  obtained by fitting to a second order polynomial in $\phi$. For the
  fit of $\phi_\mathrm{c}$, we have only considered the interval
  $\phi\in[0.52,0.55]$, while for obtaining $\phi_\mathrm{K}$, we used
  $\phi\in[0.5275,0.57]$. The two errors in the extrapolation of
  $\phi_\mathrm{c}$ correspond to the negative and positive errors.
}\label{tab:fits}
\end{table}

The quantity $\epsilon(\phi)$ carries similar information as the
configurational entropy and should go to zero at the Kauzmann transition
(see~\cite{berthier:14} for a recent similar approach). Indeed, in the
quenched case, $S_{\text con} \propto q\ \epsilon$ near the
transition. The approach followed here has the advantage that it is
ambiguity free~\cite{angelani:07}. The extrapolation at $\phi_K$
should not present problems in the quenched case (at least in the mean
field limit). In the present annealed case a more careful analysis
should be done. We can try to infer the location of
$\phi_\mathrm{K}^N$ from a second order polynomial regression (in
analogy with mean field computations~\cite{franz:98-1}), searching the
point where $\epsilon(\phi)=0$ (see Fig.~\ref{fig:epsilon} and
Table~\ref{tab:fits}). Of course, this approach leads to a very crude
estimation for $\phi_\mathrm{K}$ (the simulated values of $\phi$ are
still too far away to obtain a precise limit).  Nevertheless, our
extrapolations seem to suggest values quite smaller than the
$\phi_\mathrm{K}=0.635(2)$ obtained in~\cite{flenner:11} using
divergence of correlation times.

We have studied the equilibrium liquid-glass transition in a system of
hard spheres using a tethered Monte Carlo simulation.  This
constrained algorithm allows us to directly compute the replica
potential common in the mean field analytic computations. Using the
same formalism, we are able to present clear evidences of the
existence of a  first order transition line in the presence of
an attractive coupling between the replicas. In addition, we have
investigated the critical point, showing that it
belongs to the Ising model universality class as mean field
calculations predicted.  The emerging picture of this study is that
real glass formers seem to reproduce the same schematic phase diagram
as much simpler models.

We would like to thank in particular S. Franz and V. Mart\'in-Mayor
for interesting discussions.  The research leading to these results
has received funding from the European Research Council under the
European Union's Seventh Framework Program (FP7/2007-2013)/ERC Grant
Agreement No. [247328].

\bibliography{/home/seoane/Dropbox/biblio/biblionew}

\bibliographystyle{apsrev4-1}

\end{document}